\documentclass[twocolumn]{aastex631}
\usepackage{amsmath}
\usepackage{amssymb}
\usepackage{bm}
\usepackage{hyperref}
\usepackage{cleveref}
\usepackage{booktabs}
\usepackage{xspace}

\newcommand{\difftau}{\texttt{Diff-$\tau$}}
\DeclareMathOperator*{\argmin}{arg\,min}

\newcommand{\taurex}{\texttt{TauREx3}}
\newcommand{\viretrieval}{\texttt{VI-retrieval}}
\newcommand{\nsretrieval}{\texttt{NS-retrieval}}
\newcommand{\comments}[1]{\textcolor{black}{#1}}
\newcommand{\followup}[1]{\textcolor{black}{#1}}
\shortauthors{Yip et al.}
\graphicspath{{./}{figures/}}

\begin{document}

\title{To Sample or Not To Sample:\\ Retrieving Exoplanetary Spectra with Variational Inference and Normalising Flows}

\correspondingauthor{K.H. Yip}
\email{kai.hou.yip@ucl.ac.uk}

\author[0000-0002-9616-1524]{Kai Hou Yip}
\affil{Department of Physics and Astronomy \\
		University College London \\
		Gower Street,WC1E 6BT London, United Kingdom}

\author[0000-0001-6516-4493]{Quentin Changeat}
\affiliation{Department of Physics and Astronomy \\
		University College London \\
		Gower Street,WC1E 6BT London, United Kingdom}
\affiliation{European Space Agency (ESA), \\ ESA Office, Space Telescope Science Institute (STScI), \\ 3700 San Martin Drive, Baltimore MD 21218, USA}
\author[0000-0003-2241-5330]{Ahmed Al-Refaie}
\affiliation{Department of Physics and Astronomy \\
		University College London \\
		Gower Street,WC1E 6BT London, United Kingdom}
		
\author[0000-0002-4205-5267]{Ingo P. Waldmann}
\affiliation{Department of Physics and Astronomy \\
		University College London \\
		Gower Street,WC1E 6BT London, United Kingdom}



\begin{abstract}
Current endeavours in exoplanet characterisation rely on atmospheric retrieval to quantify crucial physical properties of remote exoplanets from observations. However, the scalability and efficiency of said technique are under strain with increasing spectroscopic resolution and forward model complexity. The situation becomes more acute with the recent launch of the James Webb Space Telescope and other upcoming missions. Recent advances in Machine Learning provide optimisation-based Variational Inference as an alternative approach to perform approximate Bayesian Posterior Inference. In this investigation we developed a Normalising Flow-based neural network, combined with our newly developed differentiable forward model, \difftau{}, to perform Bayesian Inference in the context of atmospheric retrievals. Using examples from real and simulated spectroscopic data, we demonstrate the advantages of our proposed framework: 1) Training our neural network \comments{does not require a large pre-comupted training set and can be trained with} only a single observation; 2) It produces high-fidelity posterior distributions \comments{in excellent agreement with} sampling-based retrievals and; 3) It requires up to 75\%  \comments{fewer forward model calls} to converge to the same result. 4.) \comments{This approach allows formal Bayesian model selection.} \comments{We discuss the computational efficiencies of \difftau{} in relation to \taurex{}'s nominal forward model and provide a lesson's learned account of developing radiative transfer models in differentiable languages.} Our proposed framework contributes towards the latest development of a neural-powered atmospheric retrieval. Its flexibility and \comments{significant reduction in forward model calls required for convergence,} holds the potential to \comments{be an important addition to the retrieval tool box for large and complex data sets along with} sampling-based approaches.
\end{abstract}



\section{Introduction}
Atmospheric retrieval has become an indispensable tool for astronomers to explain individual observations from transit, eclipse and phase curves spectroscopy in both low \citep[e.g.][and references therein]{Tinetti2007, Line_2013, Kreidberg_2014, Line_2014, Lee_2014, Haynes_Wasp33b_spectrum_em,Evans_2016, Tsiaras_2016, Line_2016, Tsiaras_2016_55cnc, Sheppard_2017, Stevenson_2017, MacDo_2017, Kreidberg_2018, Tsiaras2019,Evans_2019, MacDo_2019,Skaf_2020,Pluriel_2020,Zhang_2020,Changeat_2020, Chubb_2020, Alam_2020, Von_essen_2020, Anisman_2020, Carone_2021, Yip_2021, Edwards_2021,Changeat_Edwards_2021, Sheppard_2021, Saba_2021,Alam_2021,Mugnai_2021, Swain_2021, Changeat_w43,Foote_2022, Mansfield_2022_w77, Changeat_2022, evans_2022} \comments{and high-resolution \citep[e.g.][]{Brogi_2019,Gibson_2020,Molliere_2020,Seidel_2020,Boucher_2021,MacDonald_2022,Challener2022,Meech_2022,Rasmussen_2022,Harrington_2022}.} Over the years, the community has come up with a variety of retrieval frameworks, each coupled with different modelling assumptions and sampling techniques \citep[e.g.][]{irwin2008,Madhu_retrieval_method,Line_2013,helios,Gandhi_2019,Zhang_2019,Lothringer_2020,Min_2020,Cubillos_2021,taurex3}. As the number of spectroscopic observations increases \comments{with the advent of new space and ground-based observatories}, the community has started to look at planetary characterisation on a population level~\citep[][]{Sing_2016, Barstow_2017,Tsiaras2019,Pinhas_2019,Mansfield_2021, Roudier_2021, Changeat_fivekeys,edwards_pop}. 

At its core, atmospheric retrieval strives to find an atmospheric model that can best explain a given observation. Most contemporary retrieval frameworks formulated the inverse problem in terms of Bayesian statistics, where free parameters of the physical model are framed as random variables. The probability densities of these random variables ($\theta$) given the observed data ($D$) are collectively referred to as the posterior distribution, $p(\theta|D)$. The Bayes' Theorem provides a way to calculate the posterior distribution via the following relation:
\begin{equation}
    p(\theta|D) = \frac{p(D|\theta)p(\theta)}{p(D)}
\end{equation}
\noindent where $p(D|\theta)$ and $p(\theta)$ represent the likelihood and the prior distributions respectively. However, the denominator, $p(D)$, or the Evidence is intractable in most cases. The community has thus far relied on sampling techniques such as Markov Chain Monte Carlo (MCMC) or Nested Sampling to compute the approximate posterior distribution \citep[see ][for a recent review]{Madhusudhan_retreival_review}. 

Nevertheless, sampling-based techniques are prohibitively slow when the dimensionality of the problem and quantity of the observed data are large \citep[][]{VI_survey}. This issue will become increasingly pressing with the increased data volume originating from the recently launched James Webb Space Telescope  \citep[JWST,][]{greene_jwst_exo} and other future missions such as, Ariel \citep[][]{Tinetti_2021}, Twinkle \citep[][]{edwards_exo} and ELTs from the ground \citep[e.g.][]{Udry14}. These telescopes are designed to provide hundreds of higher resolution spectroscopic measurements with wider wavelength coverage. On the other hand, recent investigations have highlighted potential biases associated with some commonly used modelling assumptions, such as isothermal atmospheres, constant with altitude chemistry, or 1D plane parallel approximation \citep[e.g.][]{Rocchetto_2016, Changeat_2019, alfnoor,MacDonald_2020,Pluriel_2020_bias,Ih_2021}, prompting the need for more realistic treatment of the atmosphere \citep[e.g.][]{Irwin_2020,Feng_2020,Changeat_2020}. This will inevitably increase both the computational cost and complexity \citep[][]{Changeat_w43} of the forward model. The increase in both quantity of data and model complexity signals the need for alternative approaches to computing the posterior distribution.

Recent years have seen a surge in Machine Learning-based techniques being applied to many areas within exoplanetary science, from data detrending~\citep[e.g.][]{morvan,morvan_2,Krick_rf_irac,Gebhard_2020,Nikolaou_2020,Gebhard2022}, to planet detection~\citep[e.g.][]{Shallue_2018,Direct,Yu_tess,exominer} and to planet characterisation~\citep[e.g.][]{Marquez_rf,Zingales_2018,Cobb_2019,Waldmann_2019,Oreshenko_2020,Hayes2020,Yip_sensitivity,Haldemann2022,Ardevol_CNN,himes2020}. \comments{In 2022, the topic of planet characterisation has also been featured as a competition at the Neural Information Processing Systems \citep[NeurIPS,][]{Changeat_2022a,Yip_competition} conference.}

Variational Inference (VI) is a widely studied approach in the field of machine learning (ML) to provide approximate posterior distributions for a large and high dimensional data set \comments{with reduced computational demand compared to Markovian sampling approaches \citep[e.g.][]{Blei_VI, Fellows_VI_2018, shu2018amortized, mcvi, VI_survey, argelaguet2020mofa, friston2020dynamic,gpvae,zhang2021introduction,Karchev_2022,lopez2021deep,Lopez-Alvis2022}. However, variational methods require models that can provide their gradient with respect to some (input) parameters}. 
The field has recently explored different applications of differentiable physical models. Differentiable models open up the possibility to construct ``physics-aware" neural networks, a type of network that is explicitly constrained by physical laws \citep[e.g.][]{RAISSI2019686,chen2020physics,Amini_2021,viana2021survey,Cai_2021,haghighat2021physics,mario_pytorch,Cuomo2022}. For instance, \cite{Kawahara_2021} used Hamiltonian Monte Carlo (HMC), a gradient-informed Monte Carlo sampling algorithm \citep[][]{Duane_1987,NUTS}, to perform atmospheric retrieval of exoplanets on high resolution spectroscopic data. Others have also applied HMC to speed up lightcurve-fitting \citep[e.g.][]{Foreman_2021,Agol_2021}. 

Here we present the following contributions:
\begin{enumerate}
    \item We present \difftau{}, a \texttt{Tensorflow}-based fully differentiable atmospheric forward model, based on the implementation of \taurex \citep{taurex3, taurex3.1}.
    \item For the first time, we introduced Variational Inference as a more efficient alternative to perform atmospheric retrieval. 
    \item Our framework only requires a \textit{single} data instance during training time, there is no need for a large library of spectra for pre-training.
    \item We show that our framework \comments{formally} takes into account the uncertainty associated with the observation and is able to reproduce physically motivated correlations between atmospheric parameters.
    \item Our Bayesian Neural Network is capable of producing posterior distributions on par with distributions produced from sampling \comments{based approaches}. 
\end{enumerate}

\begin{figure*}[t]
\centering
\includegraphics[width=0.8\textwidth]{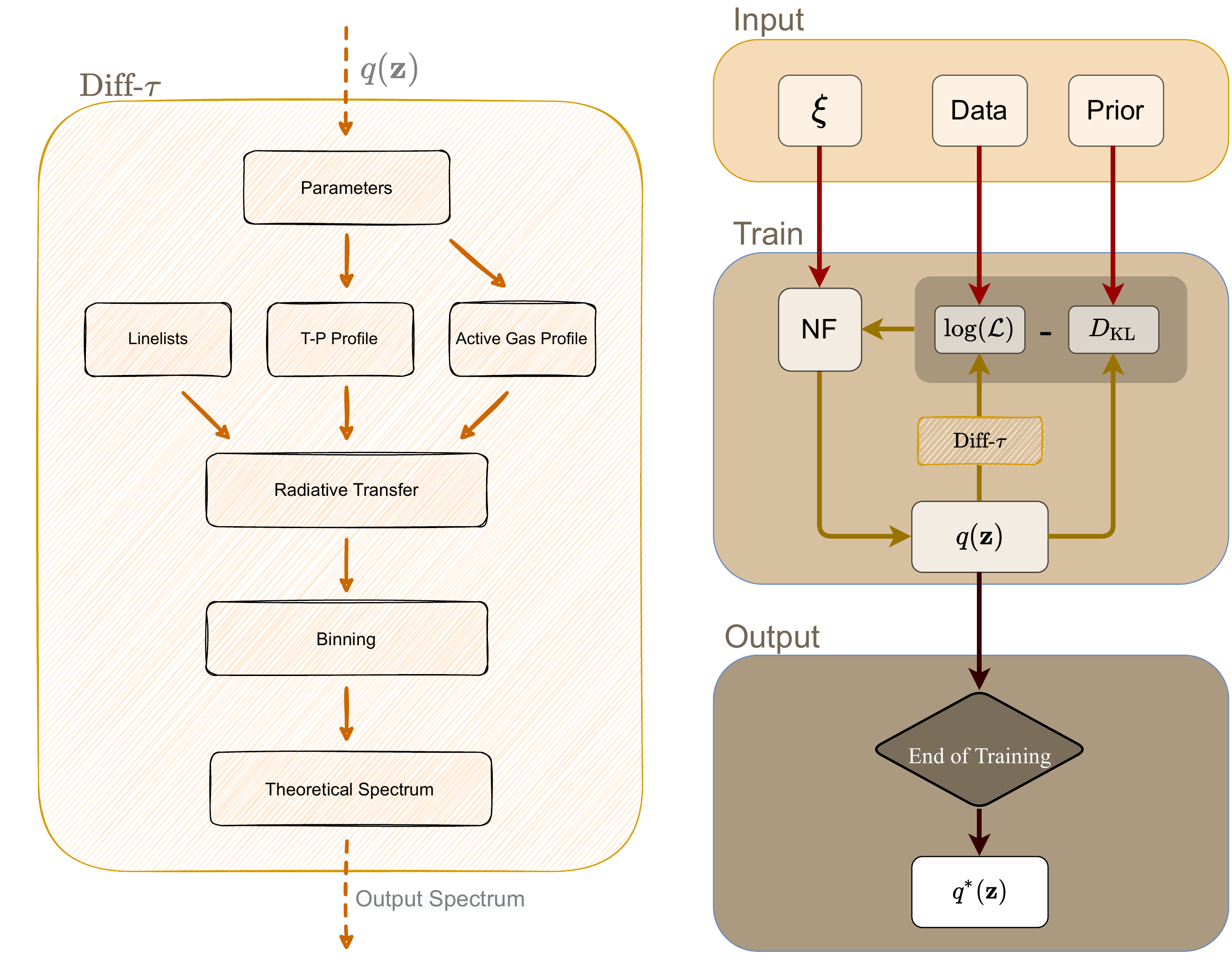}
\caption{Overview of our proposed method. Left panel shows the structure of \difftau{}. \comments{It is almost identical to a conventional forward model, in the sense that it takes in a set of physical parameters and outputs the corresponding theoretical spectrum, but with an additional ability to provide gradients.} The right panel shows the schematic of our proposed \viretrieval{}. We provided a simple (user defined) seed distribution $\bm{\xi}$, observed data and our prior bounds as inputs to the optimisation loop (Train). The Normalising Flow (NF)-based neural network, i.e. NF, is tasked to transform $\bm{\xi}$ to a surrogate distribution $q(\mathbf{z})$. We employed variational inference to approximate the typically intractable posterior distributions, where the network (NF) must learn the best transformation to optimise ELBO (controlled by log($\mathcal{L}$) and $D_{\text{KL}}$). To calculate log$(\mathcal{L})$, we used \difftau{} to transform samples from $q(\mathbf{z})$ into spectra and compared with the observed data. Once the training is completed, the trained model is able to provide a $q^*(\mathbf{z})$ that best approximate $p(\mathbf{z}|\mathbf{x})$}.
\label{fig:schematic}
\end{figure*}

\section{Overview}
In this investigation our core aim is to explore an alternative approach to conventional, sampling approach with the use of modern deep learning techniques. For simplicity we will denote conventional, sampling based approach simply as \nsretrieval{} and our proposed approach as \viretrieval{}. Our approach involves three core components - a differentiable physical model, formulation of Variational Inference and a Normalising Flow-based neural network. Here we will provide a top-level overview of how the three components interact with each other, also see \Cref{fig:schematic} for a schematic overview of the \viretrieval{}. 

Instead of relying on sampling to map the \comments{unknown} posterior distribution (as one normally does with e.g. MCMC based approaches), \viretrieval{} relies on finding a best-fit surrogate distribution to the actual posterior distribution through optimisation. The use of optimisation-based techniques means that \viretrieval{} can be orders of magnitudes faster than sampling-based approaches, especially on high-dimensional problems. However, there are two implementation difficulties that prevented the widespread use of variational methods in the field of exoplanetary atmospheres. These are: (1) \textbf{Gradients:} many optimisation procedures demand complete knowledge of the gradient flow within a computational graph (i.e. a neural network), contemporary atmospheric forward models break the flow as they are undifferentiable (2) \textbf{Surrogate distributions:} the chosen distribution (usually a multinomal Gaussian distribution) is often too simplistic to represent the actual, underlying posterior distribution.

\comments{To circumvent the above difficulties}, we built \difftau{} with the \texttt{Tensorflow} library, and utilise its automatic differentiation capabilities to compute gradients of the forward model. At the same time, we implemented a Normalising Flow-based neural network, a deep learning approach that can transform a simple, ``seed" distribution (such as a multinomial Gaussian) to arbitrarily complex distributions. 

In the following sections we will explain the theoretical background behind each technique, and in the latter part of the paper we will demonstrate how these concepts are linked to each other in practise by providing retrieval examples for three different scenarios. 


\subsection{Variational Inference}
The mathematical theory of variation inference (VI) and its application in the field of machine learning have been extensively discussed in \citep{Blei_VI}. There are on-going efforts to investigate the statistical implication of using VI for parameter estimation \citep[e.g.][]{Pati_pmlr,cherief_VI_param}. Here we provide a brief overview of the methodology. 

Given an observed spectrum $\mathbf{x}$ defined by the transit depths ($x_i$) and associate uncertainty ($\sigma_i$) at each spectral bin $i$ , the goal of an atmospheric retrieval is to find the posterior distribution $p(\mathbf{z}|\mathbf{x})$ of the set of \comments{latent} variables ($\mathbf{z}$) that can best describe the observation under a specific atmospheric model assumption $\mathcal{M}$. Instead of approximating the \comments{unnormalised} $p(\mathbf{z}|\mathbf{x})$ via sampling, variational methods approximate the distribution by finding a best-matching surrogate distribution via optimisation. 

Suppose we have a family of probability distributions $\mathbb{Q}$ parameterised by some latent variables $\mathbf{z}$. \comments{The optimal distribution (best-matching surrogate distribution to $p(\mathbf{z}|\mathbf{x})$) is one that minimises the statistical distance to $p(\mathbf{z}|\mathbf{x})$. A common choice is to compute the Kullback–Leibler (KL) divergence between the two probability density functions (\textbf{PDF})  i.e.}
\begin{align}
    q^*(\mathbf{z}) &= \argmin_{q(\mathbf{z}) \in \mathbb{Q}} \: D_{\text{KL}}[p(\mathbf{z}|\mathbf{x})||q(\mathbf{z})]\\
    &= \mathbb{E}\left[\text{log}\left(\frac{p(\mathbf{z}|\mathbf{x})}{q(\mathbf{z})}\right)\right]
    \label{eqn:KL_obj}
\end{align}
\comments{The KL-divergence measures the relative entropy between two PDFs with range [0,$\infty$]. A score of 0 means that the two distributions contain identical information, and any (positive) deviation from 0 means the two distributions become increasingly different from one another \citep[][]{kullback1951}. The KL-divergence can be computed analytically if one knows the functional form of both \textbf{PDFs}. In cases when the functional forms of one or both PDFs are unknown, as we will see below, numerical approaches must be sought to approximate the divergence. }

\comments{However, \Cref{eqn:KL_obj} itself cannot be the objective function for our optimisation task, as we do \emph{not} have any knowledge of $p(\mathbf{z}|\mathbf{x})$ to begin with}. To negate the dependence on the unknown true posterior distribution, variational inference provides an alternative formulation, the Evidence Lower BOund (ELBO), 
\begin{equation}
    \text{ELBO} = \mathbb{E}[\text{log}\,p(\mathbf{x}|\mathbf{z})] - D_{\text{KL}}[q(\mathbf{z})||p(\mathbf{z})]
    \label{eqn:elbo}
\end{equation}
where the first term is the expected value of the log-likelihood log($\mathcal{L}$) and the second term is the KL divergence between $q(\mathbf{z})$ and the prior distribution $p(\mathbf{z})$. This formulation can be understood, in terms of Bayesian Statistics, as a `tug of war' between the likelihood function and our prior belief on the distribution. We included a detailed discussion on ELBO and its link to \Cref{eqn:KL_obj} in Appendix \ref{sec:app}.

Unlike \Cref{eqn:KL_obj}, the ELBO does not require any knowledge on the intractable evidence $p(\mathbf{x})$ and therefore it can be computed analytically. We can hence find the optimal distribution $q^*(\mathbf{z})$ by minimising the ELBO. This optimisation technique is well-studied in the field of deep learning \citep{vae,IAF}. Most contemporary neural networks are trained to minimise a given loss function (e.g. ELBO in our case) by relying on a combination of gradient descent and back-propagation algorithms. Modern deep learning libraries such as \texttt{Tensorflow} \citep[][]{tensorflow2015-whitepaper}, \texttt{PyTorch} \citep[][]{pytorch} and \texttt{JAX} \citep[][]{jax2018github} provide easy access to model training and evaluation. In this study, our implementation will be solely based on the \texttt{Tensorflow} framework, but other deep learning frameworks can similarly be used \citep{Kawahara_2021}. 

\begin{figure*}
\includegraphics[width=1\textwidth]{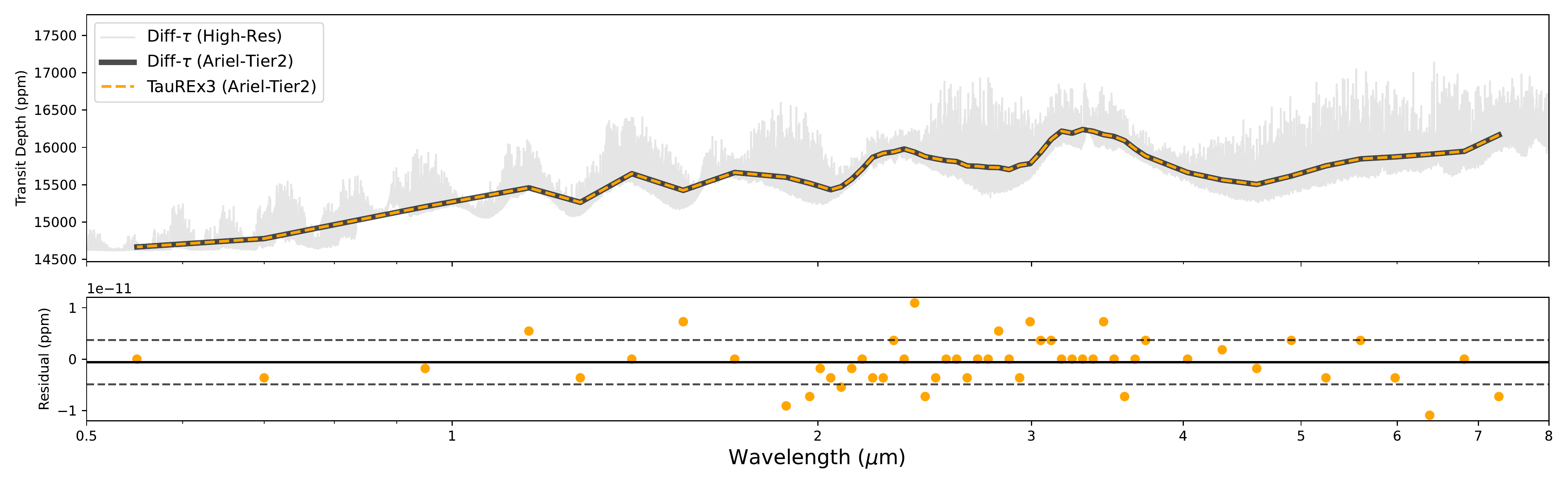}
\caption{Top panel: comparing outputs from \difftau{} in native (light grey) and binned (Ariel) resolution (orange) and \taurex{} binned to the same resolution (black dots and dashed line). The lower panel shows the residual between the two binned spectra. The grey lines represent the mean (solid line) and 1-$\sigma$ standard deviation (dashed lines).  }
\label{fig:agreement}
\end{figure*}

\subsection{Differentiable forward model, \difftau{}}
\difftau{} is an atmospheric forward model built entirely within the \texttt{Tensorflow} framework \citep[][]{tensorflow}. We followed the forward model formulation as specified in \cite{taurex3} to construct \difftau{}, with minor modifications to comply with the \texttt{Tensorflow} framework. As the code is largely based on \taurex, it provides excellent agreement between the two forward models, see \Cref{fig:agreement} for an empirical comparison between the two. We leverage the built-in automatic differentiation functionality \citep[][]{Gunes_2015}, which has the ability to automatically differentiate (almost) any functions without the need to explicitly specify the corresponding derivative form. This is immensely helpful as atmospheric models are a mixture of different physical processes, and deriving the respective derivative forms analytically can be a time consuming and non-trivial task. 


\subsection{Normalising Flows}
The stringent requirement of a pre-defined family of distributions, $\mathbb{Q}$, presents a major limitation in using VI. For many real-life scenarios, the desired posterior distributions are rarely Gaussian nor well-defined. We implemented a Normalising Flow-based neural network to break the limitations of Gaussianity in $q^*(\mathbf{z})$ in \Cref{eqn:elbo} by transforming it into an arbitrarily complex probability distribution.

Normalising Flow (NF) describes a mechanism to ``craft" a complex, multi-model distribution from a simple, ``seed" distribution~\citep[][]{NF_foundation,NF_VI,Kobyzev_2021}. This can usually be a distribution in the exponential family (e.g. a Gaussian) or an Uniform distribution. 

Suppose we have an invertible function $g$, such that we can transform a random variable $\bm{\xi} \sim p_{\bm{\xi}}$ into another random variable $\mathbf{y} \sim p_{\mathbf{y}}$ using $\mathbf{y} = g(\mathbf{\bm{\xi}})$, the probability density $p_{\mathbf{y}}$ of the random variable $\mathbf{y}$ can be computed using the change of variable formula, i.e. 
\begin{equation}
    p_{\mathbf{y}}(\mathbf{y}) = p_{\bm{\xi}}(f(\mathbf{y}))|\text{det D} (f(\mathbf{y})|
\end{equation}
where $f$ is the inverse of $g$, i.e. $f  \equiv g^{-1}$, $\text{D}f(\mathbf{y})$ is the Jacobian of $f$, i.e. $\text{D}f(\mathbf{y})= \frac{\partial f}{\partial \mathbf{y}}$ and $\text{D}g(\bm{\xi})$ is the Jacobian of $g$, $\text{D}g(\bm{\xi}) = \frac{\partial g}{\partial \bm{\xi}}$. In terms of generative models, the invertible function $g(.)$ is a generator that ``pushes" forward the seed distribution to a more complex distribution function (the generative direction). On the other hand, the function $f(.)$ moves in the opposite direction, transforming it back to a simple, ``normalised" distribution (normalising direction). It has been shown that one can generate or craft any form of distribution $p_{\mathbf{y}}$ from any base distribution $p_{\bm{\xi}}$, given that the generator $g$ can be arbitrarily complex ~\citep[][]{Bogachev_2005,Med_proof_2008}. 

Up until now we have only shifted the problem from crafting an arbitrarily complex density function to an arbitrarily complex generator function. Fortunately, invertible functions (or bijections) have a nice property - the composition of invertible functions is itself invertible, meaning that one can build a successively more complicated function by chaining non-linear invertible functions together, i.e. 
\begin{equation}
    g = g_N \circ g_{N-1} ... \circ g_{1} 
\end{equation}
where $g_{1} ... g_N $ is a set of $N$ bijective function. Similarly, $g$ has an inverse,
\begin{equation}
    f = f_1 \circ f_{2} ... \circ f_{N} 
\end{equation}
and conveniently, the determinant of the Jacobian $\text{D}f(\mathbf{y})$ is the product of individual determinant of the Jacobians $\text{D}f_i(\mathbf{y})$, i.e.
\begin{equation}
    \text{D}f(\mathbf{y}) = \prod_i^N \text{D}f_i(\mathbf{\bm{\varphi}_i})
\end{equation}
we denote $\mathbf{\bm{\varphi}_i}$ as the resultant vector of the $i$-th intermediate flow, i.e. $\mathbf{\bm{\varphi}_i} = g_i \circ \cdots \circ g_1(\bm{\xi})  = f_i \circ \cdots \circ f_N(\mathbf{y})$, where $\mathbf{\bm{\varphi}_N} = \mathbf{y}$.

\comments{In the context of our investigation, we will transform our seed distribution $\bm{\xi} \sim p_{\bm{\xi}}$ into $\mathbf{y} \sim p_{\mathbf{y}}$ and treat $\mathbf{y} \sim p_{\mathbf{y}}$ as our surrogate distribution $\mathbf{z} \sim q(\mathbf{z})$, i.e. $\mathbf{y} \sim p_{\mathbf{y}} \equiv \mathbf{z} \sim q(\mathbf{z})$}

However, these intermediate functions must -by definition- be diffeomorphic, meaning they must be bijective and differentiable (including their inverses). In recent years the field has put significant efforts in constructing bijectors that conform with these restrictions but remain sufficiently expressive and computationally efficient even in high-dimensional problems \comments{such as images \citep[][]{louizos2017,rothfuss2019,wu2020stochastic,zhang2021diffusion,survae}, notable bijector architectures include MADE \citep[][]{MADE}, Masked Autoregressive Flow \citep[][]{MAF}, NICE \citep[][]{NICE}, RealNVP \citep[][]{realnvp}, Sylvester Normalising Flow \citep[][]{berg2018sylvester}, FFJORD \citep[][]{ffjord}, Glow \citep[][]{kingma_glow_2018} and NSF \citep[][]{NSF}. Normalising Flow has proven to be a highly  successful approach in a wide range of applications, including audio synthesis \citep[][]{oord2018parallel,prenger2019waveglow,aggarwal2020}, text translation \citep[][]{jin-unsupervised,izmailov20a}, anomaly detection \citep[][]{cflow,rudolph2021}, time series forcasting \citep[][]{feng2022multi,rasul2020multivariate,schmidt2019} and image generation \citep[][]{kingma_glow_2018,ffjord,lugmayr2020srflow}. }

\section{Implementation}

\subsection{Flow-based model setup}
\label{sec:NF_setup}
We implemented the Inverse Autoregressive Flow \citep[IAF,][]{IAF} as a default bijector unit in our Normalising Flow based neural network. To perform the transformation, we chained $N=10$ bijector units together, each controlled by a 2-layer densely connected neural network with 64 hidden units and ReLU activation \citep[][]{relu} in each layer. To stabilise the network, we followed \cite{IAF} and added a Batch Normalisation layer \citep[][]{BatchNorm} after each bijector unit. We used Adam \citep{adam} as our optimiser with a scheduled learning rate (see below), the rest of the settings are kept as default from \texttt{Tensorflow}. We used a multi-dimensional uniform distribution as our seed distribution, $\bm{\xi}$.

\subsection{ELBO Formulation}
\label{sec:formulation}
We begin by describing our implementation to the general ELBO formulation in \Cref{eqn:elbo}. \comments{The ELBO consists of two terms - the log-likelihood term and the prior term. }

We defined the log-likelihood as an additive log-Gaussian PDF, i.e.
\begin{align}
    \mathbb{E}[\text{log} p(\mathbf{x}|\mathbf{z})] &= \mathbb{E}[\text{log}(\mathcal{N}(\mathbf{x}|\bm{\hat{\mathbf{\mu}}}, \bm{\sigma})] \\
     &= \mathbb{E}\left[\text{log}\left(\frac{1}{\sqrt{2\pi\bm{\sigma}^2}} \text{exp}\left(-\frac{1}{2}\frac{(\mathbf{x} - \bm{\hat{\mu}})^2}{\bm{\sigma}^2}\right)\right)\right]
    \label{eqn:llh}
\end{align}
\comments{where $\bm{\hat{\mu}}$ = $\difftau{}(\mathbf{z})$ is a forward model generated by \difftau{}} binned to the spectral resolution of the data and $\bm{\sigma}$ is the observed uncertainty. This formulation is the same as the likelihood formulation currently employed by many retrieval frameworks.

\comments{The second term, $D_{\text{KL}}[q(\mathbf{z})||p(\mathbf{z})$, computes the KL-divergence between the surrogate distribution and the prior distribution (as defined by the user), it can be expanded in a similar fashion as \Cref{eqn:KL_obj}:}
\begin{equation}
    D_{\text{KL}}[q(\mathbf{z})||p(\mathbf{z})] = \mathbb{E}\left[\text{log}\left(\frac{q(\mathbf{z})}{p(\mathbf{z})}\right)\right].
    \label{eqn:prior_term}
\end{equation}
\comments{The functional form of $p(\mathbf{z})$ can be easily determined (as it is the user defined  priori distribution). The surrogate distribtion, $q(\mathbf{z})$ has undergone multiple transformations by the Flow-based neural network. \comments{Instead of computing the KL divergence analyatical which requires knowledge of the full analytical form of the distribution}, we can approximate the KL divergence via Monte Carlo sampling, as it is not costly to sample repeatedly from the transformed surrogate distribution \citep[][]{IAF}. For this investigation, we sampled both, the prior and surrogate distributions, 10,000 times to approximate the value of the KL-divergence. Consistent with existing literature in exoplanetary retrieval, we imposed an uniform prior on all physical parameters, but we note that  any proper prior probability distribution can be used. }

Instead of optimising the ELBO objective in its actual formulation, we adopted the weight annealing approach from \cite{Sun2022} to optimise a modified ELBO objective, i.e. 
\begin{equation}
    \text{loss} = \frac{1}{\beta}\mathbb{E}[\text{log}\,p(\mathbf{x}|\mathbf{z})] - D_{\text{KL}}[q(\mathbf{z})||p(\mathbf{z})]
    \label{eqn:mod_elbo}
\end{equation}
where $\beta = \text{max}(1, \beta_0\frac{epoch}{\tau})$. $\beta_0$ and $\tau$ represent the weight constant and decay constant respectively. This formulation prevents the neural network from converging to bad local minima at the start of the training, and encourages it to explore different solutions before converging. As training progresses, the objective function will slowly converge back to the original ELBO formulation. 

\subsection{The role of \difftau{}}
The formulation of the log-likelihood function (\Cref{eqn:llh}) prompts the need to compare our input observation with theoretical spectra. The role of \difftau{} can be seen simply as a deterministic transformation, i.e. from physical parameters to spectra. As part of the optimisation process, the network must learn to adhere to physical laws imposed by the forward model. In the end, the network will produce outputs and correlations that are physically constrained in order to achieve better scores. In other words, the network's behaviour becomes physically motivated and explainable. 

\subsection{Training Procedure}
\label{sec:procedure}

\comments{At training time, we define the seed (user defined) distribution $\bm{\xi}$ = $U[\text{min},\text{max}]$\footnote{all bounded between [-1,1]}, our Flow-based neural network (the generator) will then transform it into a surrogate distribution $q(\mathbf{z})$ through our chain of bijections. We then sample from $q(\mathbf{z})$ and generate the atmospheric forward model using \difftau{}.  At each iteration, we will sample 5 times from the surrogate distribution and compute the average (modified) ELBO objective, \Cref{eqn:mod_elbo}. The entire process is repeated until the optimisation has converged or is terminated. We have implemented an \texttt{Early Stopping} procedure to stop the optimisation process if the loss value (ELBO) does not improve over 50 epochs. As we have adopted a weight annealing objective, the stopping criteria will only become effective in the later stages of the optimisation, when the modified objective converges back to the original ELBO formulation (weights stay unity)\footnote{ the weights annealing strategy reduces the contribution from the much larger likelihood function initially and slowly return it to its original value as training progresses. The loss value will almost always increase as the weights slowly increases back to unity)}. The best model is used to produce the results in Section \ref{sec:application}}. \comments{As for the learning rate, we have implemented a cyclic learning rate as suggested by \cite{smith_2015} and \cite{himes2020}. Our experiments is consistent with their claims of improved training performance and we show that it outperforms constant learning rate and a step-wise decaying learning rates\footnote{The learning rate will reduce if the loss value did not decrease after certain number of epochs.} in terms of loss value as well as speed of convergence.} 

Once trained, the generator network is decoupled from the framework. At inference time, we will initiate a seed distribution and pass it to used to transform the seed distribution into our best matching surrogate distribution.

\section{Application}
\label{sec:application}
In this section we will perform atmospheric retrieval with \nsretrieval{} and \viretrieval{}. We will demonstrate the technique through three cases:
\begin{enumerate}
    \item Real Hubble/WFC3 (HST/WFC3) observation of the hot-Jupiter HD209458\,b.
    \item Simulated Ariel Tier-2 observation of the hot-Jupiter HD209458\,b.
    \item \comments{Simulated observation (0.5\textmu m -15\textmu m) with N-point temperature profile \citep{Waldmann2015,Changeat_w43} of the hot-Jupiter WASP-43\,b.}
    
\end{enumerate}

\begin{table*}[]
\centering
\begin{tabular}{ccccccc}
\toprule
Parameters          & Ground Truth*  &     Case I           & Case II           & Priors  & Reference \\ \midrule
T$_p$ (K)           & 1449          & \checkmark            &    \checkmark      &   $\mathcal{U}$(100,2500)    & N/A \\
R$_p$ (R$_J$)         & 1.359        & \checkmark              &   \checkmark       &   $\mathcal{U}$(0.5,2.5)     & N/A \\
log$_{10}$(H$_2$O)   & -5           & \checkmark            &  \checkmark        &   $\mathcal{U}$(-12,-2)     & \cite{polyansky_h2o} \\
log$_{10}$(CH$_4$)   & -5           &             &    \checkmark      &   $\mathcal{U}$(-12,-2)        & \cite{ExoMol_CH4_new} \\
log$_{10}$(CO$_2$)   & -5           &             &   \checkmark       &   $\mathcal{U}$(-12,-2)       & \cite{Yurchenko_2020}\\
log$_{10}$(NH$_3$)   & -8           &           &  \checkmark        &   $\mathcal{U}$(-12,-2)       & \cite{ExoMol_NH3} \\
log$_{10}$(CO)      & -8           &          &  \checkmark        &      $\mathcal{U}$(-12,-2)    & \cite{Li_CO}\\
log$_{10}(\chi^{lee}_{mie}$)   & N/A          &  \checkmark              &          &   $\mathcal{U}$(-40,-4)      & \cite{lee_scattering} \\
$q^{lee}_{mie}$   & N/A          &  \checkmark             &          &   $\mathcal{U}$(1,99)    & \cite{lee_scattering} \\
log$_{10}(a^{lee}_{mie}$)   & N/A          &   \checkmark            &          &   $\mathcal{U}$(-3,1)    & \cite{lee_scattering} \\

\bottomrule
\end{tabular}
\caption{\comments{Fitted parameters for Case I and II, along with their corresponding ground truth, prior bounds, and molecular line list references for each trace gas species. The checkmark (\checkmark) indicates whether the parameter is included in each case. The ground truth is only available for Case II.}  }
\label{tab:GT_and_priors}

\end{table*}

\subsection{Atmospheric model setup for Case I and II}
\label{sec:atmo_setup}
We have taken values from \cite{Tsiaras_2016} as our reference values on the HD209458 system. In both cases, we assumed a blackbody emission for the host-star at T$_{eff}$ = 6065\,K and stellar radius R$_*$ = 1.155\,R$_{\odot}$. We also assumed a primary atmosphere (dominated by H and He, ratio = 0.175) at solar abundance. We divided the atmosphere into 70 atmospheric layers over a 10$^{-5}$\,Pa and 10$^6$\,Pa pressure range (evenly spaced in logarithmic scale) and furthermore assumed an iso-thermal T-P profile and an iso-abundance chemical profile. \Cref{tab:GT_and_priors} shows the fitted input parameters for each case and their corresponding ground truth, prior bounds and formulation/linelist references. The atmospheric setup of Case III can be found in Section \ref{sec:npts}.

\subsection{Case I: HST/WFC3 observation of HD209458\,b}
Case I aims to demonstrate our method's applicability to actual data. The transmission spectrum is observed with the HST/WFC3 G141 grism and processed by \texttt{Iraclis} \citep{Tsiaras_2016, Tsiaras2019}. The detrending process is described in details in \cite{Tsiaras_2016}. The basic atmospheric setup follows Section\,\ref{sec:atmo_setup}. For this case we are assuming a hydrogen/helium dominated atmosphere, with opacities from trace gases absorption and Mie scattering clouds\footnote{We note that the alternative and commonly used flat cloud model (i.e. a constant in pressure opacity cut-off) is inherently not differentiable and not physically viable, we therefore chose to include the differentiable Mie scattering formulation in \difftau{}.} \citep{lee_scattering}. We ran the optimisation procedure for 2000 epochs, \comments{with the convergence parameters, $\beta_0$ and $\tau$, set to 100 and 200 respectively}. These values are determined with a coarse hyperparameters search between $\beta_0$ = [100, 1000] and $\tau$ = [100, 1000].

\subsection{Case II: Ariel Tier-2 observation on HD209458\,b}
In the second case we would like to understand the ability of our proposed framework to retrieve the ground truth value and showcase the flexibility of our framework to switch to different atmospheric assumptions and spectral resolutions. We used \texttt{ArielRad} \citep[][]{mugnai_Arielrad} to simulate the expected noise level for each wavelength channel for HD209458\,b at Ariel Tier\,2 resolution. In terms of atmospheric chemistry, we adhered to the same setup as described in Section\,\ref{sec:atmo_setup} and include five trace gases H$_2$O, CH$_4$, CO and CO$_2$ and NH$_3$. We chose this set of molecules due to their expected contribution in the wavelength range considered, and because they have been successfully detected in hot Jupiter's atmospheres. \comments{We ran the optimisation procedure for 2000 epochs, with $\beta_0$ and $\tau$ set to 100 and 300 respectively. }

\subsection{Case III: N-Points TP profile retrieval with WASP-43\,b}
\label{sec:npts}
\comments{We simulated the atmosphere of a WASP43\,b like planet observed at a customised wavelength range, i.e. from 0.5 - 15 \textmu m. The atmospheric model setup largely followed Section \ref{sec:atmo_setup}, but with 50 atmospheric layers instead of 70 layers. As for stellar and planetary parameters, we followed \cite{Hellier2011} and set T$_{eff}$ = 4400\,K, R$_*$= 0.6\,R$_{\oplus}$, M$_p$ = 1.78\,M$_J$ and R$_p$ = 0.93\,R$_J$. We included log$_{10}$(H$_2$O) and log$_{10}$(CH$_4$) as trace gases in the atmosphere, with their abundances set at -4 and -3 respectively. As for the temperature-pressure profile, we adopted the dayside temperature profile of WASP-43\,b retrieved by \cite{Changeat_w43}. We then generated a (binned) atmospheric model with the above settings, and added a 100\,ppm Gaussian white noise to produce an observation. The prior bounds of the parameters follows Table \ref{tab:GT_and_priors} apart from planet temperature, where we expanded the bounds to [100, 3000]. The observation remains the same for all retrievals performed in Case III. }

\comments{As for the retrieval settings, we performed a N-Point retrieval with each temperature point located at a fixed, pre-defined pressure point. The top and bottom pressure points are fixed at $10^{-5}$ and $10^{6}$ respectively throughout the investigation, while the (pressure) points in between them are varying in accordance to the number of retrieved temperature (-pressure) points. In all cases the pressure points are separated equally (in log-pressure space). Other retrieved parameters included R$_p$, log$_{10}$(H$_2$O) and log$_{10}$(CH$_4$). In summary, the total number of free parameters is N (temperature points) + 3. The network setup mostly follows Section \ref{sec:procedure}. We ran the optimisation procedure for 5000 epochs, with $\beta_0$ and $\tau$ as 100 and 300 respectively.}

\section{Results}
\subsection{Case I and II}
\begin{figure*}
    \centering
    \includegraphics[width=0.8\textwidth]{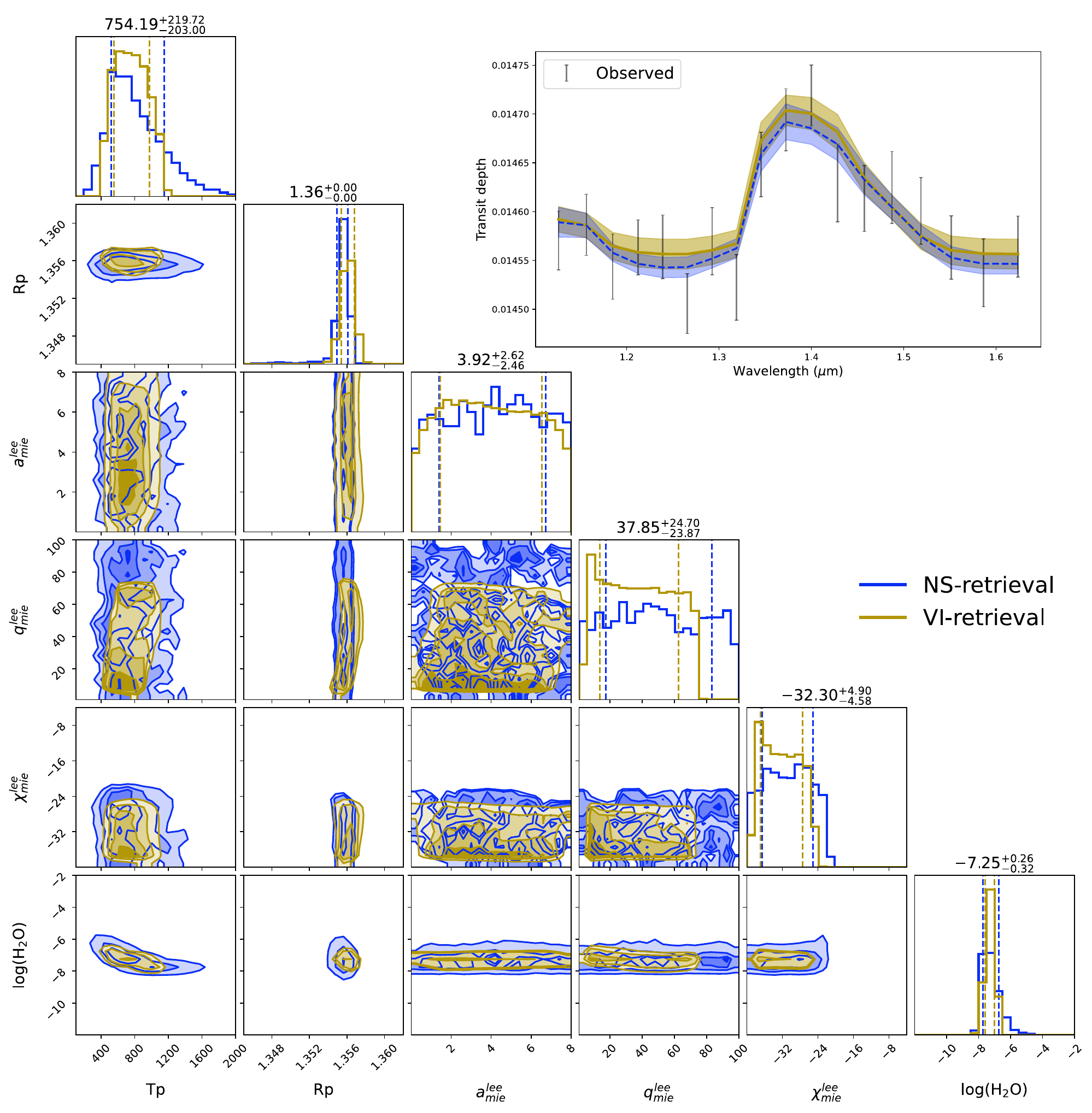}
    \caption{Posterior distribution of HD209458\,b WFC3/G141 observation obtained from \nsretrieval{} (blue) and \viretrieval{} (yellow). The top right corner is an empirical comparison between the two best fitted spectra as obtained by the two methods. The shaded area shows the 1-$\sigma$ spread of the respective retrieval approach.}
    \label{fig:hst_retrieval}
\end{figure*}
\begin{figure*}
    \centering
    \includegraphics[width=0.8\textwidth]{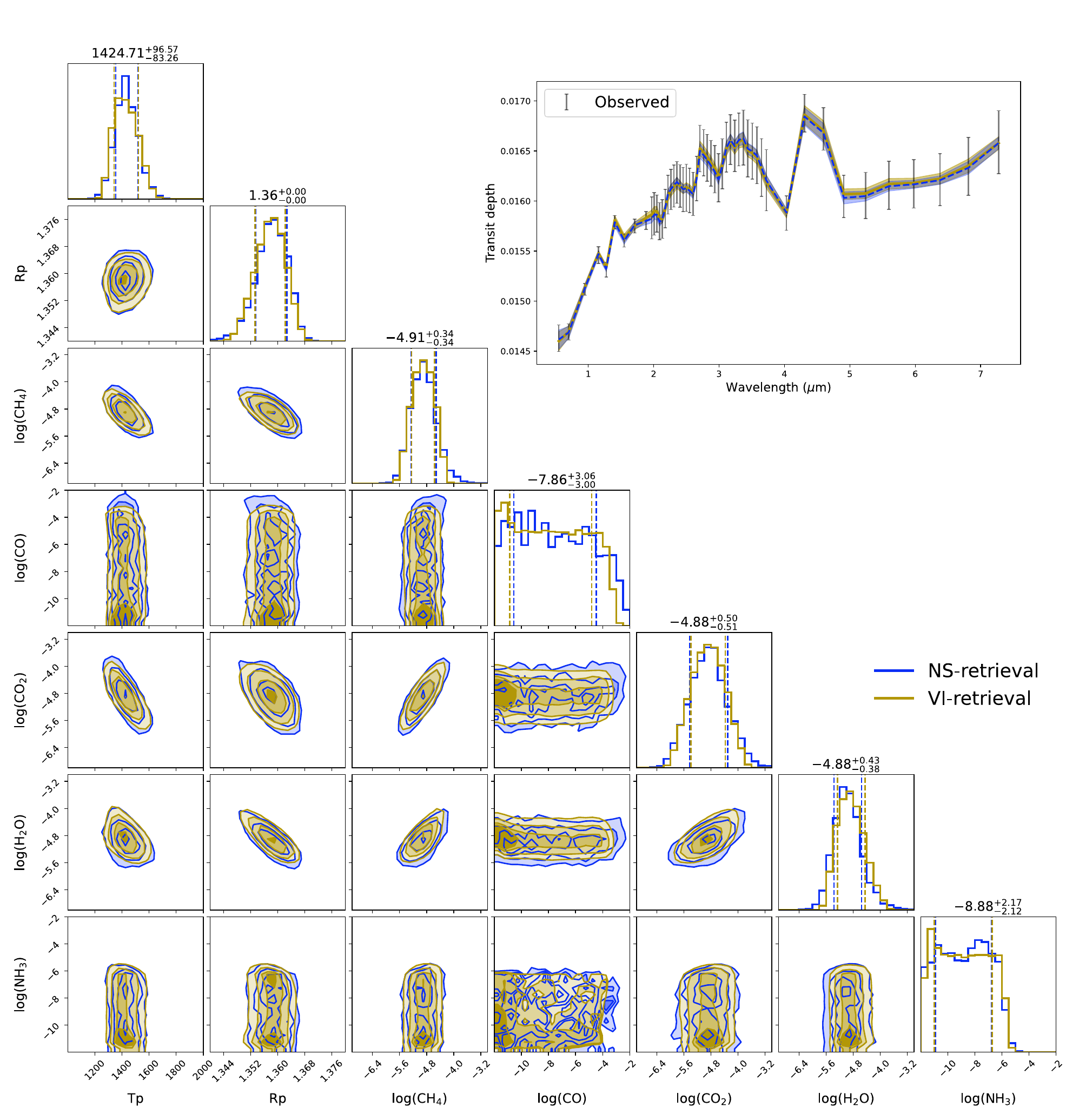}
    \caption{Posterior distribution of HD209458\,b in Ariel Tier2 resolution obtained via \nsretrieval{} (blue) and \viretrieval{} (yellow). The top right corner is an empirical comparison between the two best fitted spectra as obtained by the two methods. The shaded area shows the 1-$\sigma$ uncertainty of the respective retrieved methods}
    \label{fig:ariel_retrieval}
\end{figure*}

\comments{In both cases (Figures~\ref{fig:hst_retrieval} \& \ref{fig:ariel_retrieval}), \viretrieval{} (in yellow) is able to converge to very similar results to \nsretrieval{} (in red).}\comments{ We observed an excellent agreement between both methods and the percentile values}\footnote{\comments{defined as the 16th and 84th percentile, indicated by the dashed line in figures~\ref{fig:hst_retrieval} \& \ref{fig:ariel_retrieval}}} are within 1-$\sigma$ of each other. 

As for the shape of the posterior distribution, the surrogate distribution is able to reproduce peaked, Gaussian-like distributions whenever there is sufficient constrains from the observation (i.e. radius of the planet at 1\,bar pressure), and ,at the same time, produce a uniform-like distribution (with upper limit) when the free parameters are unconstrained by the observation (e.g. CO in Case II), conforming with our uniform prior. In addition to this, the surrogate distribution provides a faithful reproduction of the covariances between the free parameters. These correlations are resulted from the interactions between the free parameters through the formulation of the radiative transfer equations. They are most notable in Case II, when the high S/N observation of Ariel allows better constraints on the free model parameters. 

Apart from the similarities, there are also noticeable differences between the two retrievals, for instance \viretrieval{} is less adept at capturing ``cliff" like distributions (e.g. log$_{10}(\chi^{lee}_{mie}$)), when there is a sharp upper or lower bound on a parameter. There are also instances (such as T$_p$ in Case I) when the surrogate distribution is only able to capture ``part" of the conditional distribution obtained via \nsretrieval{}. The difference could be due to the imperfect optimisation process. The loss function (ELBO) is a balance between the log-liklihood term and the prior term. Minimising one term will always come at the expense of maximising the other. This situation is aggravated by the fact that the variability of the log-likelihood term is orders of magnitude higher than the prior term, which means that the training will always be driven by the former term and produce sharply peaked distributions. Our modified objective function (\Cref{eqn:mod_elbo}) aims to alleviate the imbalance between the two terms by explicitly lowering the contribution from the log-likelihood term during the start of the training. However, the underdispersed situation will likely return as the formulation slowly converge back to the original ELBO form, as seen in both cases. An alternative remedy is to increase permanently the contribution from the prior term, which will inevitably increase the estimated uncertainty and makes it harder to compare to the conventional approaches (i.e. \nsretrieval{}). 


\subsection{Case III}
\begin{figure*}[h]
    \centering
    \includegraphics[width=0.8\textwidth]{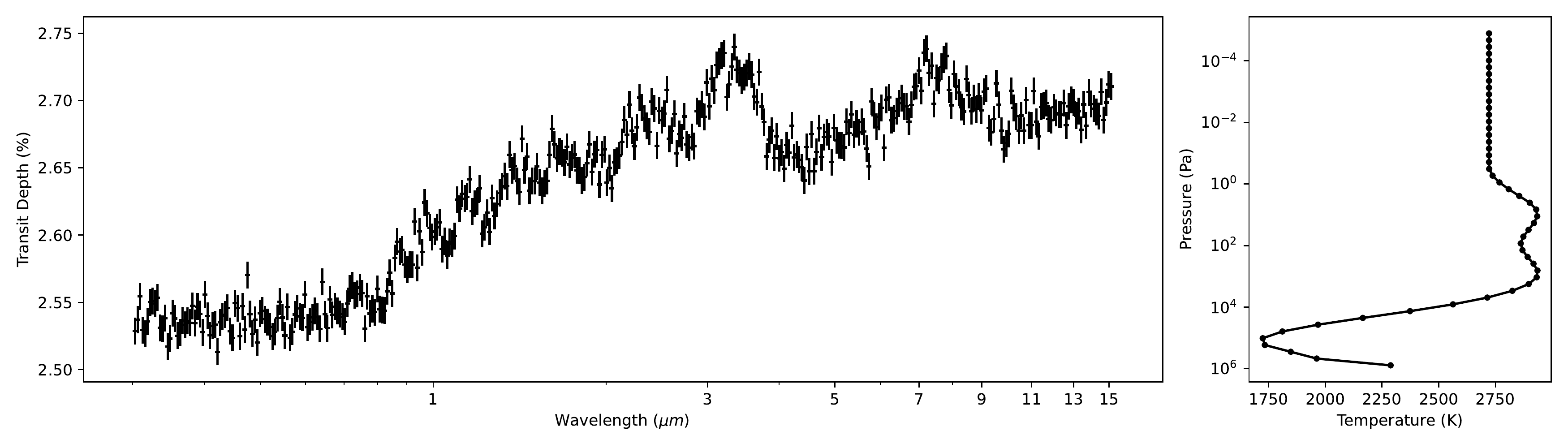}
    \caption{Left: Simulated observation of WASP-43\,b from 0.5 \textmu m to 15 \textmu m; Right: Dayside temperature profile used to generated this observation (adopted from \cite{Changeat_w43}), there are 50 pressure points  separated evenly (in logarithmic scale) between $10^{-5}$ to $10^{6}$.}
    \label{fig:npt_w43}
\end{figure*}

\begin{figure*}
    \centering
    \includegraphics[width=0.9\textwidth]{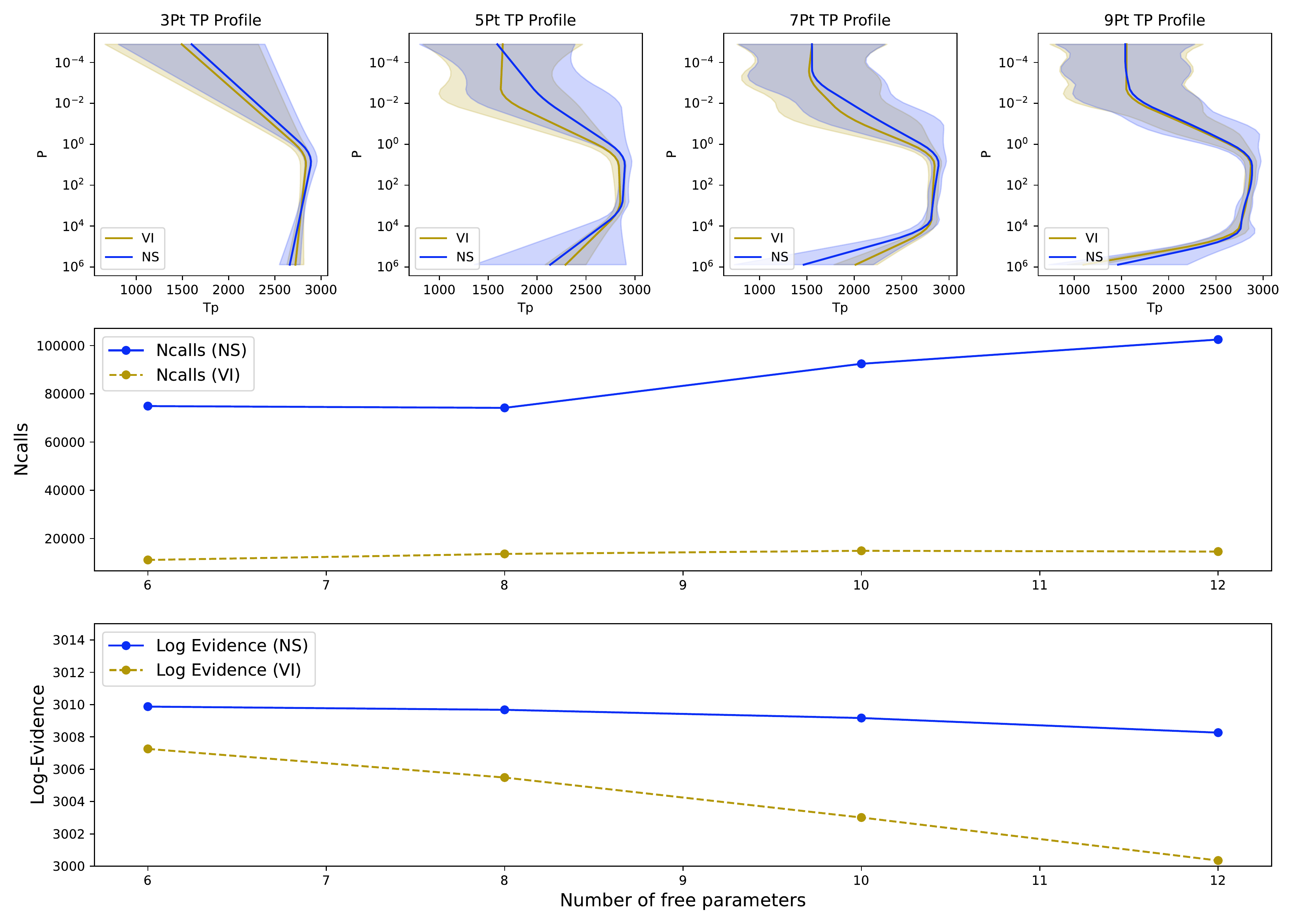}
    \caption{Comparing the performance of \viretrieval{} (yellow) and \nsretrieval{}(blue). First row: T-P profiles retrieved by each approach, each subplot shows a different fixed-point TP-profile, going from 3 points to 9 points at a step of 2. Second row: The number of forward model calls requires for each N-points retrieval. Third row: The log-evidence values retrieved by each approach. }
    \label{fig:npt}
\end{figure*}

Figure \ref{fig:npt_w43} shows the outcome of the simulated WASP-43\,b observation. Figure \ref{fig:npt} shows the results of performing N-points retrievals from 6 free parameters to 12 free parameters. The first row compares the retrieved TP profiles from both approaches (blue: \nsretrieval{}, yellow: \viretrieval{}). The second row compares the number of forward models calls (Ncalls) required by each approach. In all cases \viretrieval{} (in yellow) requires 6 times \textit{fewer} Ncalls compared to \nsretrieval{} (blue). The third row compares the log-evidence obtained by each approaches.  In all cases, the log-evidence obtained by \viretrieval{} never goes higher than the ones obtained via \nsretrieval{}, which is consistent with the formulation of ELBO. Both approaches show a declining trend in log-evidence as the number of retrieved points increases, with \viretrieval{} declining more rapidly than \nsretrieval{}. The faster decline in log-evidence may be affected by the increase in dimensionality (free parameters), which makes the accurate estimation of the log-evidence harder.  

\section{Discussion}
\subsection{Limitations of Grid-based learning}
\comments{Most contemporary machine learning based atmospheric retrievals  are trained in a supervised fashion with a large grid of simulated spectra produced by a forward model. We broadly refer these models as grid-based models here. This kind of training procedure takes away the computational burden of having to generate thousands to millions of forward model on the fly during model deployment and makes these models computationally fast for applications within their trained domain. In other words, these models offload the computational burden of atmospheric retrievals to the model training stage, with the fully trained model being quick to run at inference.
However, this approach comes with three limitations:  1) model generalisability, 2.) lack of Bayesian framework and 3.) a lack of model interpretability. }


\begin{table}[]
\centering
\begin{tabular}{rcc}
\toprule
Ncalls                & \nsretrieval{} & \viretrieval{} \\ \midrule
WFC3/G141       & 105622      &  8781    \\
Ariel           & 65414       & 9570 \\
\bottomrule
\end{tabular}
\label{tab:Ncalls}
\caption{Comparing the number of forward model calls by each method for the same atmospheric models (as in Case I) in WFC3/G141 and Ariel spectral resolution. In both cases, the number of forward model calls by \viretrieval{} is significantly lower than \nsretrieval{}.}
\end{table}

\paragraph{Generalisability}
\comments{Any changes to the underlying model, spectral range or resolution, will hinder the model's performance, and in some cases, will require a full re-computation of the training data from scratch and re-training the model ~\citep{Marquez_rf,Zingales_2018,Cobb_2019,Yip_sensitivity,Haldemann2022,Ardevol_CNN}. Such a scenario can be triggered by anything as simple as adding an extra molecule that is previously not present in the training data.
These limitations can be alleviated to a certain extent by training a surrogate forward model as done in \cite{himes2020}. Their model is tasked to produce synthetic spectra at very high resolution, and the output can subsequently be down-sampled to any appropriate spectral range and resolution when required. Nevertheless, the model is not immune to changes to the underlying atmospheric assumptions and will likely need to be re-trained in those cases.}


\paragraph{Lack of Bayesian framework}
\comments{Most contemporary retrievals aim to map the Bayesian posterior distribution. In contrast, most ML models applied in the field of atmospheric characterisation are formulated to perform maximum likelihood estimation~\citep[e.g.][]{Marquez_rf,Yip_sensitivity,Haldemann2022,Ardevol_CNN}. The difference between these two objectives  presents an obstacle when trying to compare outputs from the two methodologies. Other non-Bayesian approaches includes \texttt{ExoGAN}, in which case the generator is trained to optimise the adversarial loss \citep[][]{Zingales_2018}, but this loss function does not guarantee a robust computation of $P(\theta |D)$. There are other ways to bypass this limitation. For instance,  \cite{Cobb_2019} and \cite{Ardevol_CNN} use an ensemble of neural network to approximate the conditional distribution. \cite{himes2020} bypassed this constraint by running a traditional retrieval with a ML-based surrogate forward model.}

\paragraph{Interpretability}
\comments{Interpretability varies from one model to another. Learning algorithms such as linear regression and Decision trees are some of the most transparent  algorithms, but with limited modelling capability. Deep learning algorithms sits at the opposite end of the spectrum, with extensive learning capabilities, but at the cost of very limited interpretability. Most grid-based learning models ~\citep{Marquez_rf,Zingales_2018,Cobb_2019,Yip_sensitivity,Haldemann2022,Ardevol_CNN} used more complex models to learn the covariance between parameters from a large grid of spectral examples. However, the loss of interpretability means that it becomes hard for users to understand when, where and how the algorithm may break. With no explicit control on the learning process (remember they are only asked to optimise the learning objective), it is uncertain how the model may interpret the training data. By relying only on the training data, it becomes important to make sure the training data can adequately present the underlying forward model, which may be computational expensive or in some cases difficult to ascertain \citep{Fisher2022}. }

 
 

\subsection{A Flexible and Interpretable Bayesian Framework}
 
 
  \comments{Motivated by the above limitations. We presented an alternative approach to train an end-to-end deep learning model for atmospheric retrievals. Instead of focusing on a generalisable model (one that works on a wide range of spectra), our framework is specific (or in other words, overfitted) to the observed data, and similar to conventional Markovian sampling algorithms, it is not generalisable and must be re-run for any changes in observed data and/or model assumptions. By dropping the goal of training a generalisable model, we forego the need to create a vast training sets as well as the need to pre-train our model before any retrieval could take place. It furthermore affords us with the flexibility of easily changing our input data and forward models. This flexibility has allowed us to explore the performance of our model at different spectral resolutions, wavelength ranges, observational uncertainties and model assumptions with relative ease (as demonstrated above). However, this alternative approach also comes with disadvantages, see Section \ref{sec:limit}. } 
  
  \comments{To allow direct comparison with conventional retrievals, we formulated the objective function to optimise the ELBO function, an alternative formulation of the Bayes' Theorem. This modification constrains the behaviour of our trained model, and allows us to obtain results comparable to conventional retrievals (as demonstrated above). Another advantage of this approach is the drastic reduction of the number of forward model computation to a fraction of its retrieval counterpart. Table \ref{tab:Ncalls} shows a comparison of the number of forward model calls between \texttt{NS-} and \viretrieval{} in Case I and II.}
  
 
 \comments{In the grid-based approaches, models have to learn implicitly the physical relationships between atmospheric parameters from the training data. There is no guarantee that physical laws are preserved or correctly represented in the trained model. Our framework is purposely built to explicitly impose physical laws. One can view our NF-based network as a generator and \difftau{} as a corresponding decoder that transforms parameters into spectra. The static\footnote{static in the sense that it remains invariant throughout the training} decoder acts as a physical regularizer to the generator, which explicitly constrains the output from the generator to align with physical laws (any misalignment will be reflected in the transformed spectra). This is demonstrated through our examples, the surrogate distributions are able to provide physically plausible correlations and are aligned with correlations produced from standard Bayesian Sampling retrievals (\nsretrieval{}, blue contour). Having a fully analytic model allows us to impose physical laws directly, without going through a training data proxy. }

\subsection{Objective Function}
The objective (loss) function is a crucial factor that governs the learning behaviour of a deep learning model. Deep learning models in the literature are usually trained to best-match the respective ground truth values of the physical parameters\footnote{parameter values used to generate the forward model}(fitting-for-parameters). Here we opt to align our objective to that of our retrieval counterpart, in other words, we are explicitly asking our model to look for solutions that can best explain our observation (fitting-for-observations). 

\comments{Adopting an observation based likelihood not only allows us to align with the objective function of conventional retrievals, it also has added advantage of properly accounting for observational uncertainty. ML-based retrieval methods often incorporate observational uncertainties through noise augmentation \citep[][]{Yip_sensitivity, Ardevol_CNN}, this has resulted in over-estimation of the error bound as compared to Nested Sampling based retrievals \citep[][]{Ardevol_CNN}. }

\comments{In an ideal world, both approaches (fitting-for-parameters and fitting-for-observations) will agree with each other. However, it is not the case for inverse problems, where our observation is inherently corrupted\footnote{possible sources include instrument and astrophysical noise sources as well as the information loss from binning data and/or forward model} and we may never be able to recover the ground truth in some cases due to the loss of information and the inverse processing being ill-defined. In such cases, differences in the objective function may lead to different results. On one hand, the fitting-for-parameter approach is asking the neural network to pursue parameter values that may no longer be possible to retrieve (due to corruption of the observed data), which will cause the neural network to exhibit fictitious behaviour if it results in the lowering of the loss function values \citep[][]{Yip_sensitivity}. On the other hand, the fitting-for-observation approach explicitly ask for spectra that can explain the observation, \textit{not} the underlying ground truth. Of course, that will also mean that our approach is not immune from the intrinsic retrieval biases induced by the atmospheric forward itself \citep[][]{Rocchetto_2016,Feng_2020, MacDonald_2020}. These biases can only be alleviated through increasing the complexity of the atmospheric forward model to better represent the physical/chemical processes leading to the observed spectra.}



In terms of model development, our framework bypasses the need to train the network with a large library of synthetic spectra before applying to actual data, as we are training the network directly on actual observations. This move avoids the problem of data shift \citep[][]{datashift} - where our training distribution is different from our test distribution.

\subsubsection{Computational cost of \viretrieval{}}
\comments{In this section we will discuss the computational cost associated with running a retrieval using \viretrieval{} compared to \nsretrieval{}. Under the same forward model, the \viretrieval{} will always beat the \nsretrieval{}, but the matter becomes slightly more complicated as the two methods do not always reuse the same forward model.  We provide an initial assessment of the two methodologies with two different forward models. Note that the following figures are based on running both methods using a Macbook Pro with Apple M1 chip (8 CPU cores in total). These figures should be taken with caution as the two forward models are at different developmental stages, TauREx3 is a highly optimised code while \difftau{} is proof-of-concept forward model written from scratch entirely in Tensorflow}.

\comments{A single forward model call (\texttt{FMcall}) for Case I scenario takes about $ 0.89 \pm 0.02 $ seconds on \difftau{} and $0.26 \pm 0.10 $ seconds with TauREx3's forward model. For a single retrieval, \viretrieval{} takes 8781 \texttt{FMcall} and \nsretrieval{} takes 105,622 \texttt{FMcall}. Based on these numbers, \viretrieval{} takes 7815s and \nsretrieval{} takes 94,003s respectively. As for Case II, a single forward model call takes about $ 5.08 \pm 0.38 $ seconds on \difftau{} and $0.51 \pm 0.34 $ seconds with TauREx3's highly optimised forward model. For a single retrieval, \viretrieval{} takes 9,570 \texttt{FMcall} and \nsretrieval{} takes 65,414 \texttt{FMcall}. Based on these numbers, \viretrieval{} takes 48,615.2s and \nsretrieval{} takes 33,361s respectively.}


From our simple analysis above we can see that \viretrieval{}'s computation time is dependent on the speed of the forward model. It is empirically faster than \nsretrieval{} if the computational time of the forward models are similar to each other (Case I), but the advantage will elapse as the forward model takes longer to compute . This difference in performance can be easily narrowed down with model optimisation and maturity of the framework itself. In this vein, \viretrieval{}'s will play an important role in retrieval scenarios with complex, and subsequently slow to run, forward models. In these situations the optimisation objective becomes the number of forward model calls with \viretrieval{}'s being an order of magnitude more efficient.

\subsubsection{Model Selection}
\label{sec:selection}

\begin{table}[]
\centering
\begin{tabular}{rcccc}
\toprule
Model             & ELBO & Ref  & log$_{10}(\mathcal{B})$\\
\midrule
Flat line         & 62.74        & 62.83 & 315.66 \\
No Methane        & 345.37        & 347.18 & 33.03 \\
Complete          & 378.40        &  380.20 & N/A\\
Overspecified Model & 374.00     & 377.74 & 4.4\\
\bottomrule

\end{tabular}

\caption{Retrieved ELBO from \viretrieval{} and Bayesian (log-) Evidence from \nsretrieval{} (Ref) for each scenario. log$_{10}(\mathcal{B})$ shows the log-difference between the ELBO of our complete model and other competing models. \label{tab:model_selection}}
\end{table}

Model selection is a key part of model evaluation cycle. So far, the machine learning retrieval literature has largely omitted the issue by training networks under one or several fixed atmospheric assumptions. Our flexible framework and similar objective function to \nsretrieval{} means that we can, for the first time, utilise some of the tools frequently used by sampling-based retrievals to compare the retrieval results from different models. 

Given that our surrogate distribution is a good approximation of the underlying posterior distribution, our ELBO, despite being a lower bound, should closely approximate the Bayesian evidence. We can therefore use the ELBO as a proxy of the evidence, and compare our models by estimating the Bayes' factor. 
\begin{equation}
    \mathcal{B}_{10} = \frac{P(\textbf{x}|\mathcal{M}_1)P(\mathcal{M}_1)}{P(\textbf{x}|\mathcal{M}_0)P(\mathcal{M}_0)}
\label{eqn:bayesfactor}
\end{equation}
where P$(\textbf{x}|\mathcal{M}_k)$ represents the Bayesian Evidence attained from model $\mathcal{M}_k$, $P(\mathcal{M}_k)$ represents our prior belief on a particular model. 

As an empirical example, we took our example from Case II and performed \viretrieval{} and \nsretrieval{} with different atmospheric assumptions, including a flat line model, an incomplete model (without methane), a complete model (as specified in Case II) and an overspecified model (Case II + TiO and VO). Table \ref{tab:model_selection} compares the corresponding ELBO from \viretrieval{} and Bayesian Evidence from \nsretrieval{}. The ELBO retrieved from each model closely follows, but always smaller than, the corresponding Bayesian Evidence, as set out from the definition of ELBO. The Bayes' factor displayed in Table \ref{tab:model_selection} allows us to differentiate different models. Following Jeffrey's guideline scale \citep[][]{jeffreys_theory_1998, Hobson2002, Padilla_2019}, the Bayes' factor strongly favours our complete model over other competing models. For more discussion on using Bayesian evidence as a model selection tool, please refer to the Appendix in \cite{Changeat_w43}.

\subsection{Limitations}
\subsubsection{Limitations on \viretrieval{}}
\label{sec:limit}
In this section we will focus on the framework of \viretrieval{}. 
\begin{enumerate}
    \item \textbf{Generalisbility:} As set out from the design goal, \viretrieval{} is targeted to work on a single observation. This approach has earned us flexibility, where we can freely change our model assumptions, spectral range and resolution with relative ease. However, the same advantage has also limited the generalisbility of our framework. As opposed to other deep learning approaches, which can be rapidly deployed to a data set within its training set range, it will have to be re-trained for each observation.
    \item \textbf{Convergence:} Gradient descent helps to converge to an global optimum if function is convex. However in the presence of model degeneracy, the function becomes non-convex and convergence to global minima is not guaranteed. 
    \item \textbf{Flow Bijections:} The expressiveness of the surrogate distribution depends highly on the architecture of neural network and the type of Flow bijectors (IAF in our case). While our investigation has demonstrated the flexibility if these bijectors, it does not mean that \textit{any} distribution \citep{durkan2019neural} can be mimicked. It remains an ongoing effort to design bijectors that are both flexible and computational efficient. 
    \item \textbf{Approximation:} The formulation of variational inference, especially the use of ELBO in the objective function, means that the retrieved surrogate distribution will always be an approximation to the ground truth. The use of Flow-based neural networks has of course improved the fidelity of the approximation.
    \item \textbf{Tendency to produce underdispersed solution:} \followup{The implementation of ELBO and its modified form in Section \ref{sec:formulation} hints the tendency for the network to produce underdispersed solution. The negative log-liklihood term (1st term) is almost always going to be larger than the prior term (2nd term), meaning that the network will tend to produce a sharp distribution first before widening itself to comply with the prior. This effect is also related to the expressiveness of the bijections, their flexibility is ultimately bounded by the range of `action' that they can perform to transform the distribution. We note here that this effect is minor in our case as the shape of the target distributions are relatively simple. }
    \item \textbf{Hyperparameter tuning:} Similar to conventional neural networks, hyperparameters do not always stay optimal from one setup to another, which means some degree of hyperparameter search should be performed to identify a good setup However, this is alleviated by identifying the key hyperparameters that may influence the optimization procedure, in our case we realised the annealing weights plays a role in the optimisation procedure, owing to its close relationship with the objective function. Other hyperparameters are generally fixed and do not differ much.  
\end{enumerate}

\subsubsection{Limitations on differentiable frameworks}
\label{sec:comments}
The ability to differentiate and provide gradients with respect to some quantities is central to modern deep learning algorithms. The growing popularity of Machine Learning in recent years has accelerated the development of differentiable frameworks, among them \texttt{Tensorflow}, \texttt{PyTorch} and \texttt{JAX} have amassed a substantial user bases. 

There have been several recent attempts within the field of exoplanets to develop differentiable physical models within these frameworks. Initial results show that these differentiable models may hold the keys to overcome the curse of dimensionality brought by our increasingly complex models \citep[][]{mario_pytorch,Kawahara_2021} and may one day enable us to perform population study without placing significant demand on computational resources.

However, these frameworks are not without issues. Here we would like to provide our ``user experience" for readers who are interested and/or would like to implement their own models. 

\begin{enumerate}
    \item \textbf{Significant overhead:} Current development in differentiable programming mandates that any implementation must be written entirely in terms of a chosen framework. While it has become relatively straightforward to translate \emph{most} operations from one framework to another, it is not a trivial and error-free process. These frameworks come with their own programming restrictions and conventions. Users are expected to adhere to these conventions or otherwise they might risk losing the ability to differentiate.
    \item \textbf{Differentiability is not guaranteed:} Not all operations are differentiable. \comments{Some operations may be mathematically non-differentiable\footnote{such as a discontinuity in the function, e.g. logarithmic function} or they are not designed to have a gradient, such as a \texttt{LookUpTable} or \texttt{interpolation}. Depending on the algorithmic structure of the forward (atmospheric) model, this obstacle may present difficulties in obtaining a valid gradient. One good example is the grey cloud model. This model assumes the planet becomes completely opaque below a certain pressure level/altitude. This model, while easy to implement in any computational language, is not differentiable. A possible way around is to implement their own gradient like in \cite{Kawahara_2021}, but this approach is only possible if one knows the differentiable form.} 
    \item \textbf{Sub-optimal Performance:} Many modern differentiable languages are built and designed around deep learning based applications. They are not designed to handle complex computational models. In other words, a differentiable model may suffer from reduced performance compared to its un-differentiable counterpart \citep[][]{hu2019difftaichi,hu2019taichi}. 
    \item \textbf{Rapidly evolving language:} Differentiable languages are under constant development and rapid release cycles in response to latest researches. Some of these developments may not be backward-compatible\footnote{one good example is the switch from \texttt{Tensorflow} 1 to \texttt{Tensorflow} 2.} and may impact the long term sustainability of the developed model. 
\end{enumerate}

\section{Conclusion}

In this paper we introduced the differentiable forward model (\difftau{}) based on \taurex{} and implemented in \texttt{Tensorflow}. We combined our newly developed model with our density-alternating neural network and showed that it is possible to compute the approximate Bayesian posterior distribution \comments{that are in excellent agreement  with the ones produces from computationally more expensive sampling based techniques. }
Through our examples we have demonstrated three advantages of our framework:
\begin{enumerate}
    \item \textbf{Fewer forward model calls}: Our \viretrieval{} requires 25\% or fewer forward model calls compared to \nsretrieval{} to converge, which opens up opportunities for more rapid retrieval with more complex (i.e. slower) forward models. 
    \item \textbf{Flexibility}: Unlike many deep learning based frameworks, which rely on a large pre-computed libraries of spectra, our proposed framework resembles more closely a traditional retrieval set up by explicitly including the physical forward model in the deep learning architecture. Consequently, it retains many of the advantages of a traditional retrieval codes, such as the freedom to choose the model assumptions, number of free parameters, spectral wavelength range and observed uncertainty, without having to produce a separate training dataset each time.
    \item \comments{\textbf{Error propagation}: By incorporating observational uncertainty directly into the likelihood function, we demonstrated the capability of our network to produce uncertainty bounds that are on par with conventional atmospheric retrievals. }
    \item \textbf{Model Selection}: We demonstrated, for the first time, that we can compare the adequacy of our neural network model to a given observation by computing the Bayesian Evidence and Bayes Factor. 

\end{enumerate}

Our proposed framework presents a major step towards the wider adoption of neural network powered atmospheric retrievals. With significantly higher spectral resolutions and signal-to-noise ratios afforded by JWST and Ariel data, atmospheric forward models must consequently increase in complexity to model these data accurately. Such an increase in the dimensionality of the problem will result in significant strain on traditional sample-based retrieval approaches.  
\comments{While the results shines light to rapid AI-assisted Bayesian inference, we must stress that our framework is \emph{not} meant to replace sampling based framework, but to \emph{complement} existing frameworks. \viretrieval{}, similar to most MCMC based approach, are susceptible to non-global minima, \nsretrieval{} on the other hand, tends to allow more thorough exploration of the parameter space. This difference becomes important when multiple solutions are equally plausible for a given observation.} The approximate retrieval performed by \viretrieval{} can act as a precursor before the more computationally heavy Nested Sampling retrievals are required. Its \comments{rapid convergence} means that one can scan through multiple candidate atmospheric models before converging into a few promising models for further, more detailed evaluations. The parameter bounds extracted by our framework can also act as an informative prior to speed up the sampling process of conventional retrievals.

This project has received funding from the European Research Council (ERC) under the European Union’s Horizon 2020 research and innovation programme (grant agreement No 758892, ExoAI) and the European Union’s Horizon 2020 COMPET programme (grant agreement No 776403, ExoplANETS A). Furthermore, we acknowledge funding by the Science and Technology Funding Council (STFC) grants: ST/K502406/1, ST/P000282/1, ST/P002153/1 and ST/S002634/1. 


%

\vspace{5mm}


\software{TauREx 3~\citep[][]{taurex3, taurex3.1}, Tensorflow2 ~\citep[][]{tensorflow2015-whitepaper}, tensorflow-probability ~\citep[][]{dillon2017tensorflow}, 
scipy~\citep[][]{2020SciPy-NMeth}, NumPy~\citep[][]{harris2020array}, Pandas ~\citep[][]{reback2020pandas},nestle~\citep[][]{barbary_2021}, corner~\citep[][]{corner}, matplotlib~\citep[][]{matplotlib}, h5py~\citep[][]{collette_python_hdf5_2014}
          }



\appendix

\section{ELBO derivation}
\label{sec:app}

In this section we will describe the mathematical derivation that led to the formulation of the Evidence Lower BOund, ELBO. We will start from \Cref{eqn:KL_obj}, i.e. 
\begin{equation}
    q^*(\mathbf{z}) = \argmin_{q(\mathbf{z}) \in \mathbb{Q}} \: D_{\text{KL}}[p(\mathbf{z}|\mathbf{x})||q(\mathbf{z})]
\end{equation}
We start by expanding the left hand side of the above expression:
\begin{equation}
    D_{\text{KL}}[p(\mathbf{z}|\mathbf{x})||q(\mathbf{z})] = \mathbb{E}[\text{log}q(\mathbf{z})] - \mathbb{E}[\text{log}p(\mathbf{z}|\mathbf{x})] \\
\end{equation}
and expand the latter term with the multiplication rule and chain rule of probability, 
\begin{align}
    D_{\text{KL}}[p(\mathbf{z}|\mathbf{x})||q(\mathbf{z})] &= \mathbb{E}[\text{log}(q(\mathbf{z}))] - \mathbb{E}[\text{log}p(\mathbf{z},\mathbf{x})] +\text{log}p(\mathbf{x}) \\
     &= \mathbb{E}[\text{log}(q(\mathbf{z}))] - \mathbb{E}[\text{log}p(\mathbf{x}|\mathbf{z})] + \mathbb{E}[\text{log}p(\mathbf{z})]+ \text{log}p(\mathbf{x}) \\
     &= -(\mathbb{E}[\text{log}p(\mathbf{x}|\mathbf{z})] - (\mathbb{E}[\text{log}(q(\mathbf{z}))] - \mathbb{E}[\text{log}p(\mathbf{z})]) + \text{log}p(\mathbf{x}) \\
     &= -(\mathbb{E}[\text{log}p(\mathbf{x}|\mathbf{z})] - D_{\text{KL}}[q(\mathbf{z}) || p(\mathbf{z})]) + \text{log}p(\mathbf{x}). \\
\end{align}
Here we will substitute the expression for ELBO as defined in \Cref{eqn:elbo},
\begin{equation}
    D_{\text{KL}}[p(\mathbf{z}|\mathbf{x})||q(\mathbf{z})] = - \text{ELBO} + \text{log}p(\mathbf{x})
\end{equation}
From the above expression we can see that the KL divergence is determined by the interaction between the ELBO and the Bayesian Evidence term. While the former term depends on the quality of the surrogate distribution $q(\mathbf{z})$, the latter term is a constant as it depends only on the data $\mathbf{x}$. This, combined with the fact that $D_{\text{KL}} \geq 0$ , the ELBO term is inherently constrained by $\text{log}p(\mathbf{x})$, i.e.
\begin{equation}
    \text{ELBO} \leq \text{log}p(\mathbf{x})
\end{equation}
Hence the term is known as the Evidence Lower BOund (ELBO). We can therefore use ELBO as our objective function, since maximising the ELBO is equivalent to minimising the KL divergence. In our implementation, the objective function is set to minimise the negative ELBO. 

\bibliography{sample631}{}
\bibliographystyle{aasjournal}



\end{document}